\documentclass{scrartcl}
\usepackage{graphicx,hyperref,authblk,booktabs}
\usepackage[margin=1in]{geometry}

\begin{document}

\title{Visualization of Diseases at Risk\\
	in the COVID-19 Literature}	
\author{Francis Wolinski}
\affil{Yotta Conseil, France\\
\texttt{\small{\{forename.surname\}@yotta-conseil.fr}}}
\date{}

\maketitle

\begin{abstract}
	This paper presents a project, named VIDAR-19, able to extract automatically diseases from the CORD-19 dataset, and also diseases which might be considered as risk factors. The project relies on the ICD-11 classification of diseases maintained by the WHO. This nomenclature is used as a data source of the extraction mechanism, and also as the repository for the results. Developed for the COVID-19, the project has the ability to extract diseases at risk and to calculate relevant indicators. The outcome of the project is presented in a dashboard which enables the user to explore graphically diseases at risk which are put back in the classification hierarchy. Beyond the COVID-19, VIDAR has much broader applications and might be directly used for any corpus dealing with other pathologies.
\end{abstract}

\section{Introduction}

With the spread of the novel coronavirus, named \emph{SARS-CoV-2}, and the corresponding disease \emph{COVID-19} -- which contaminated more than 3.4 million people and killed more than 240 thousands people as of 2 May 2020 \cite{worldometers} -- many initiatives have been launched for powering research to control the epidemic.

One of the cross-cutting research priorities of the World Health Organization research roadmap deals with vulnerable population subgroups. They include ``those suffering with stigmatization, the elderly, those with co-morbidities and the immunocompromised'' \cite{whoroadmap}.

The fact is that reports on risk factors and comorbidities are spread in many publications and do not provide an analytic view with actionable information\cite{science}. Being able to collect automatically such risk factors from literature and to provide a visualization of their importance is at stake.

We developed the project \textbf{VIDAR-19} (\textbf{VI}sualization of \textbf{D}iseases \textbf{A}t \textbf{R}isk in CORD-\textbf{19}) to cope with this problem. It is able to extract automatically from the CORD-19 dataset \cite{cord19} all diseases referenced in the International Classification of Diseases (ICD-11) maintained by the WHO \cite{whoicd11}, as well as the diseases which might be considered as risk factors.

Along with the extractions, the process calculates several indicators which are made available at different levels of the ICD-11 classification. The results are showcased in a dashboard. It enables the user to explore and to visualize graphically in an intuitive and convenient way the diseases at risk which are put back in the classification hierarchy.\\

\section{Materials and Methods}

In this project, the data sources are:

\begin{itemize}
	\item the COVID-19 Open Research Dataset (CORD-19) including scientific papers from 5 different sources \cite{cord19};
	\item the International Classification of Diseases ICD-11 available on the WHO web site \cite{whoicd11}.
\end{itemize}

The overall project is performed in 3 steps:

\begin{itemize}
	\item The first step loads and prepares the ICD-11 dataset and produces a repository able to store the classification as well as the results of the extraction step. See details in subsection \ref{loading}.

	\item The second step implements a mechanism for extracting from texts disease names of ICD-11, risk factors, and additional filters as well. See details in subsection \ref{extraction}.

	\item The third step calculates several indicators and provides meaningful visualizations on the diseases that have been found in the corpus and those which might be considered as risk factors. See details in subsection \ref{visualization}.
\end{itemize}

This work has been implemented using the Python language and a few libraries of its Data Science and Natural Language Processing ecosystem.

\subsection{Loading and preparation of ICD-11 classification}\label{loading}

The ICD-11 classification is available in a spreadsheet collected from the WHO web site \cite{whoicd11}. The diseases in ICD-11 are organized in a classification hierarchy. For instance, the disease \emph{Pneumonia} is classified as this:

\begin{center}
\emph{Diseases of the respiratory system} \textgreater \emph{Lung infections} \textgreater \emph{Pneumonia}
\end{center}

\subsubsection{Loading of ICD-11}

Two columns of this dataset are of interest in this project. The column \emph{Code} stands for a unique disease identifier, and the column \emph{Title} stands for the name of diseases. An extract of the raw dataset for the disease \emph{Pneumonia} and its branch is shown below \ref{icd11}.

\begin{figure}[h]
	\centering
	\begin{tabular}{lll}
	\toprule
	{} &  Code &                               Title \\
	\midrule
	6856 &   NaN &  Diseases of the respiratory system \\
	7002 &   NaN &                   - Lung infections \\
	7003 &  CA40 &                       - - Pneumonia \\
	\bottomrule
\end{tabular}

	\caption{Extract of the raw ICD-11 (\emph{Pneumonia} branch)}
	\label{icd11}
\end{figure}

We can notice that in this dataset, the depth of the classification hierarchy is represented with a tabulation prefix in the \emph{Title} column of the spreadsheet, and materialized by a sequence of dashes. We can also notice that upper level branches of the classification do not hold any code (\emph{NaN} values).

\subsubsection{Preparation of ICD-11}

To achieve our goal, we need to enhance this basic information into a higher level representation of the classification. The resulting dataframe object is organized with a column for each level in the classification hierarchy, so that any disease is along with its full path from the root to itself.

If \emph{Harry Potter} holds a wand, the \emph{Data Scientist} has \emph{pandas} \cite{pandas}. In that case, there is no need for latinized spells, but a clear and powerful API, which verbs that are obviously magic.

\begin{figure}[h]
	\centering
	\resizebox{\textwidth}{!}{
		\begin{tabular}{lllrlllllllll}
\toprule
{} &  Code &                               Title &  Level &                                   0 &                1 &          2 & 3 & 4 & 5 & 6 & 7 & 8 \\
\midrule
6856 &   NaN &  Diseases of the respiratory system &      0 &  Diseases of the respiratory system &                  &            &   &   &   &   &   &   \\
7002 &   NaN &                     Lung infections &      1 &  Diseases of the respiratory system &  Lung infections &            &   &   &   &   &   &   \\
7003 &  CA40 &                           Pneumonia &      2 &  Diseases of the respiratory system &  Lung infections &  Pneumonia &   &   &   &   &   &   \\
\bottomrule
\end{tabular}
}
	\caption{Extract of the prepared ICD-11 (\emph{Pneumonia} branch)}
	\label{full_df}
\end{figure}

With this representation, any piece of information attached to a disease with a code -- in columns added by a further process -- will be also accessible from its upper levels in the classification hierarchy.

\subsection{Extraction of ICD-11 diseases and risk factors in COVID-19 literature}\label{extraction}

\subsubsection{Preparation of the extraction}

For this purpose, the \emph{flashtext} \cite{flashtext} library is suitable to deal with complex disease names such as those contained in the ICD-11. This library implements a powerful regular expression engine, named \emph{keyword processor} able to search for phrasal keywords in any text and in a single pass.

In this project, 3 \emph{keyword processor} instances are built:

\begin{itemize}
	\item A first instance is initialized from ICD-11 disease names with a code (15,286 diseases in our case, since we do not consider entries after the branch \emph{External causes of morbidity or mortality} of the classification, and we discard a few single-word entries which are sources of ambiguities). However some disease names need to be preprocessed prior entering them into the keyword processor:

\begin{itemize}
	\item Some names are often compound of 2 parts separated by a coma. For instance, the disease: \emph{Coronavirus infection, unspecified site}. In such cases, the rightmost part is an additional information. It is discarded prior to feed the keyword processor in order to better match the literature which seldom mentions the full disease name.
	\item Some names are often not expressed in literature by using the official title used in the ICD-11. For instance, the common noun \emph{cancer} is more likely used than the official term \emph{carcinoma}. Here again, in order to better match the literature, those disease names are automatically preprocessed to feed the keyword processor with synonyms. For instance, \emph{carcinoma of breast, specialised type} is firstly transformed into \emph{carcinoma of breast} (see above), and then completed automatically with the additional expression \emph{breast cancer}.
	\item In addition, synonyms of diseases have also been used. They are accessible from the ICD API made available by the WHO \cite{icdapi}.
\end{itemize}

\item A second instance is initialized from a list of expressions representing risk factors or comorbidities, such as: \emph{risk factor}, \emph{high-risk factor}, \emph{morbidity factor}, \emph{comorbidity factor}, etc. A list of 40 expressions, including plural forms, dealing with risk factors has been established by using a bigram finder from the \emph{nltk} library along with the \emph{Word2Vec} algorithm from the \emph{gensim} library, to extend semi-manually the list with synonyms. This task is not detailed in the present paper.

\item A third instance is initialized from a list of expressions representing additional filters of documents. In this project, we defined a single filter to tag the documents dealing explicitly with the COVID-19 (disease or virus) by using alternatives expressions such as: \emph{SARS-CoV-2}, \emph{COVID-2019}, \emph{2019-nCoV}, etc. When activated, this filter selects 6,760 documents from the whole corpus.

\end{itemize}

In all cases, the idea is to build a comprehensive list of triggering expressions so as to maximize the recall of the extracting/filtering process.

\subsubsection{Processing of the COVID-19 corpus}

The COVID-19 Open Research Dataset (CORD-19) is a growing resource of scientific papers on COVID-19 and related historical coronavirus research \cite{cord19}.

In this project, for each document 5 pieces of information are directly collected: the data source it belongs to, the document id, and also the title, abstract and body text. The processing of a document extracts 3 additional information thanks to the 3 keyword processor instances defined above:

\begin{itemize}
	\item unique ICD-11 codes of diseases which are mentioned in the whole document;
	\item unique ICD-11 codes of diseases which are mentioned along with risk factors or comorbidities in the document\footnote{The granularity is a paragraph: either the title, or abstract paragraphs, or again body text paragraphs.};
	\item a boolean whether or not the document deals with the COVID-19 (disease or virus).
\end{itemize}

These information are organized by data source and by document, so as to produce later the visualizations for the whole document repository, or for a selection of data sources, or again for the documents selected by activating the additional filter. Moreover, this representation enables to make incremental updates of the repository by adding continuously upcoming documents.

As of 2 May 2020, 48,410 documents have been processed (46,106 documents with a disease and 5,153 documents with risk factors) with 15,286 diseases from ICD-11 (3,370 diseases found in the corpus and 1,423 diseases with risk factors).

\section{Results}

\subsection{Calculation of indicators}

For the visualization, several indicators are calculated on 3 sets of diseases along with the classification hierarchy: diseases with a code in ICD-11 ($S_{0}$); diseases found in the corpus ($S_{1}$); diseases at risk ($S_{2}$). The calculations can be performed for the whole ICD-11 classification hierarchy, or for a branch, and also for the whole corpus, or again for a subset of documents.

Let $C$ be a considered set of documents from the corpus, $S$ be a set of diseases, $B$ a branch (or a sub-branch) of diseases, $d$ a disease and $n(d)$ the number of documents including the disease.\\

We define:

\begin{itemize}
	\item Share of a branch: $\frac{\vert{B}\vert}{\vert{S}\vert}$ (applied to: $S_{0}, S_{1}, S_{2}$), see \ref{shares};
	\item Occurrences of diseases in a branch: $\sum_{d \in B} n(d)$ (applied to: $S_{1}, S_{2}$), see \ref{occurrences};
	\item Document frequency of a disease: $\frac{n(d)}{\vert{C}\vert}$ (applied to: $S_{1}, S_{2}$), see \ref{frequencies}.
\end{itemize}

\subsection{Visualizations}\label{visualization}

The visualizations are produced with a few graphical libraries: \emph{matplolib} \cite{matplotlib} for stacked bar charts, its extension \emph{seaborn} for bar charts, and also \emph{plotly} \cite{plotly} for treemap graphics (and also funnel graphics not presented in this paper).

For the share and occurrences of branches, the user has the ability to select a specific branch to explore, in order to figure out the over- or under-representation of some diseases.

\subsubsection{Shares of branches}\label{shares}

A stacked bar chart shows the shares of branches for the 3 sets of diseases defined above. In the figure below, the chart is produced for the whole ICD-11 classification \ref{shares_icd11}. We notice an over-representation of the shares for several branches: e.g., \emph{Certain infectious or parasitic diseases} (green, left), \emph{Diseases of the immune system} (gray, left), \emph{Endocrine, nutritional or metabolic diseases} (cyan, left), \emph{Diseases of the circulatory system} (pink, middle), \emph{Diseases of the respiratory system} (blue, middle), \emph{Symptoms, signs or clinical findings, not elsewhere classified} (red, right).

This chart can also be produced for a specific branch, below \emph{Endocrine, nutritional or metabolic diseases} \ref{shares_obesity}. We notice an over-representation of the shares for: \emph{Endocrine diseases} (blue) and \emph{Nutritional disorders} (blue).

\subsubsection{Occurrences of diseases by branch}\label{occurrences}

The following 2 treemap graphics are based on the occurrences of the 2 sets of diseases defined above (in the corpus vs. at risk) by branch. These graphics provide some fisheye views on the ICD-11 classification hierarchy, with high level of detail in the focus areas \cite{treemap}, which is in our case related to the number of occurrences.

These graphics can be produced for the whole ICD-11 classification \ref{treemap1} \ref{treemap2}. We notice in the second treemap graphics an over-representation of several branches: e.g., \emph{Diseases of the respiratory system} (blue), \emph{Endocrine, nutritional or metabolic diseases} (cyan), \emph{Diseases of the circulatory system} (pink).

\begin{figure}[h]
	\centering
	\includegraphics[width=5.5in]{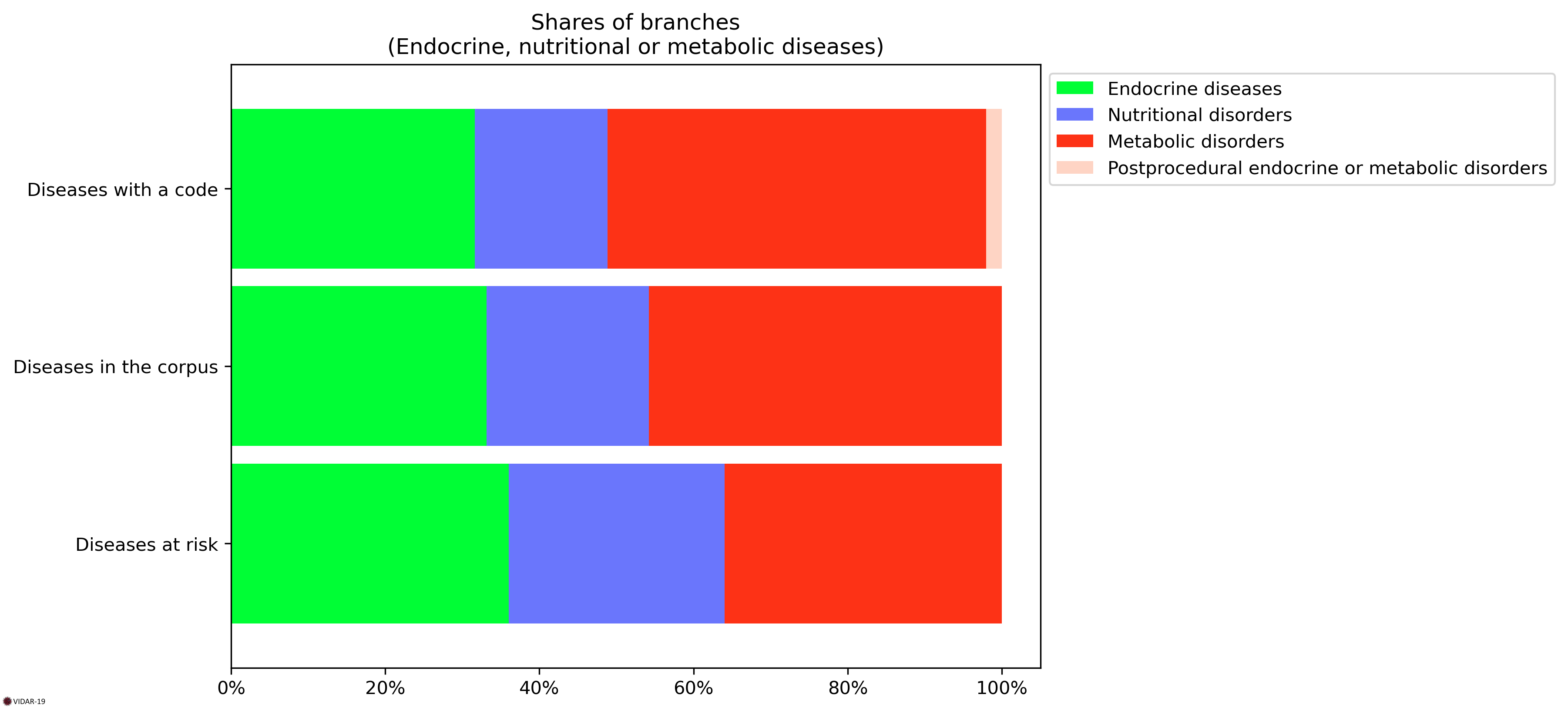}
	\caption{Comparison of the shares of diseases for the \emph{Endocrine, nutritional or metabolic diseases} branch}
	\label{shares_obesity}
\end{figure}

These treemap graphics can also be produced for a specific branch, here \emph{Endocrine, nutritional or metabolic diseases} \ref{shares_obesity}. We notice an over-representation for: \emph{Obesity} (blue) and \emph{Diabetes} (green).

\subsubsection{Document frequency of diseases}\label{frequencies}

Beyond these meaningful graphics, it is possible to compare the document frequency of the diseases which have the highest gap between the 2 sets of diseases defined above (at risk vs. in the corpus). Below, we present the chart obtained when activating the additional filter for documents dealing explicitly with the COVID-19 (disease or virus) \ref{gaps}.

We compare in a later section the list of diseases mentioned in this chart with the lists of diseases at risk that have been already published for the COVID-19, see \ref{discussion}.

\subsection{Aggregating the results into a dashboard}

\begin{figure}[h]
	\centering
	\includegraphics[width=5.5in]{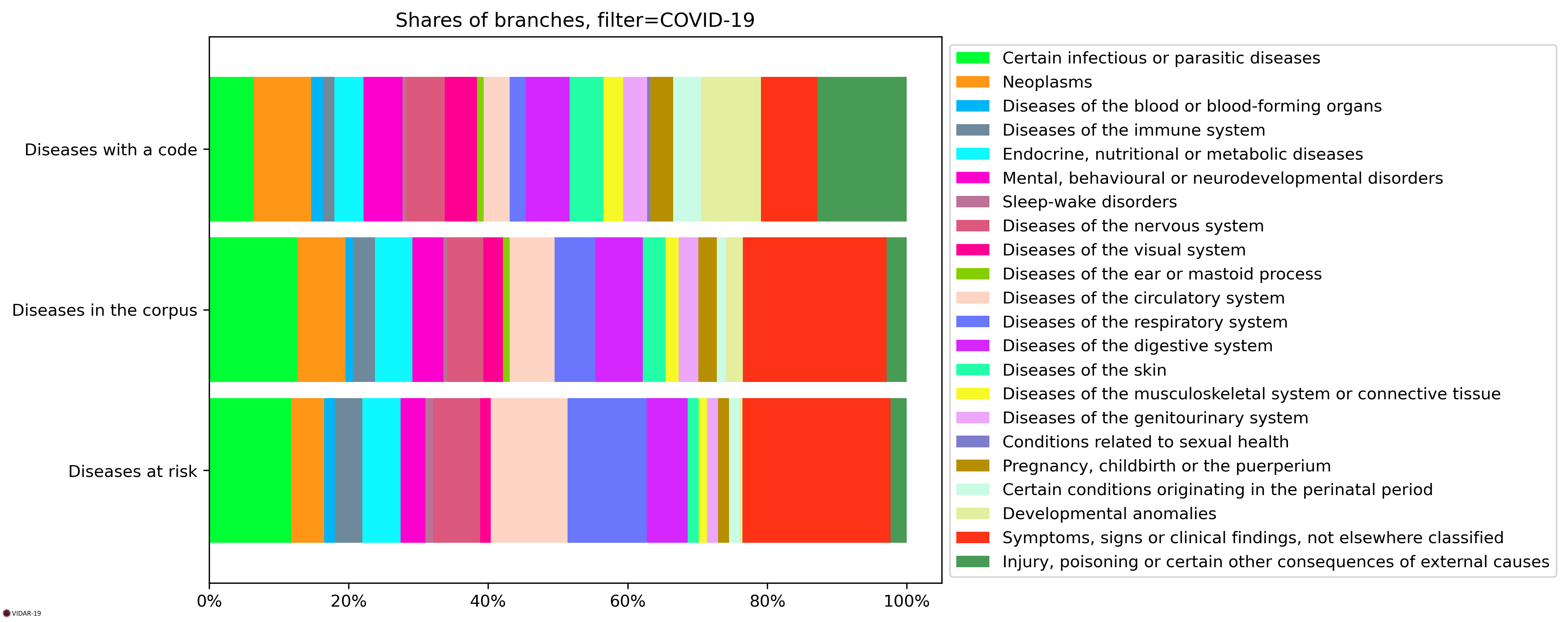}
	\caption{Comparison of the shares of branches for the whole ICD-11 classification}
	\label{shares_icd11}
\end{figure}

\begin{figure}[h]
	\centering
	\includegraphics[width=5.5in]{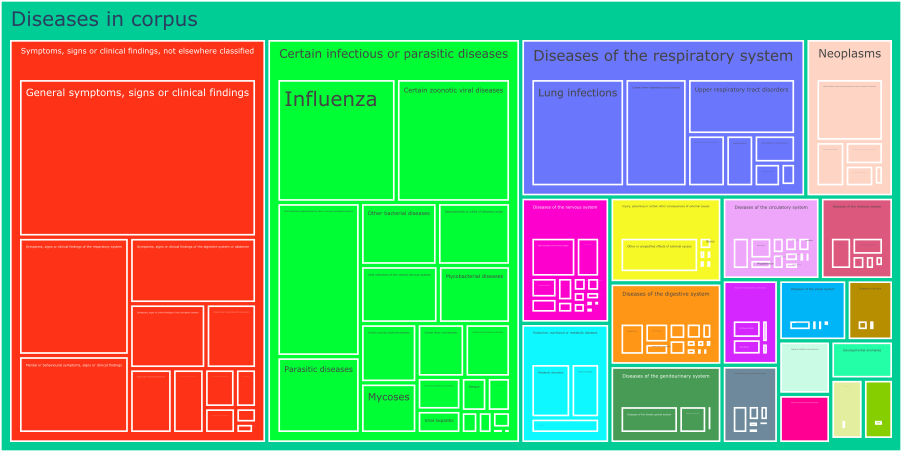}
	\caption{Occurrences of diseases in the corpus (all branches)}
	\label{treemap1}
	\vspace{0.5in}
	\includegraphics[width=5.5in]{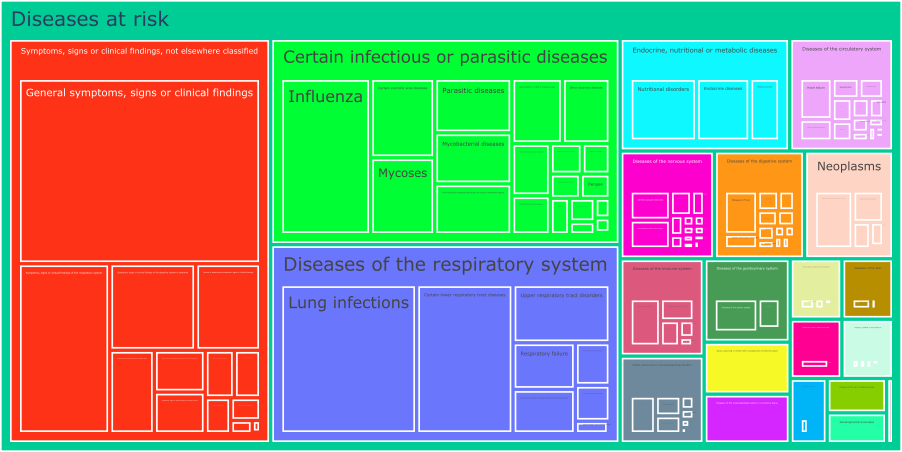}
	\caption{Occurrences of diseases at risk (all branches)}
	\label{treemap2}
\end{figure}

In order to ease the exploration and the exploitation of the results by a user, we have built a data driven dashboard able to showcase this project. To do so, some ergonomics constraints and utilities have been added to its design and go beyond the results presented above:

\begin{itemize}
	\item The order of diseases, or branches, in the charts with shares \ref{shares_icd11} fits the breadth-first traversal of the ICD-11 classification;
	\item The colors of diseases, or branches, in the charts with shares \ref{shares_icd11}, and those used in the treemap graphics with occurrences \ref{treemap1} \ref{treemap2} are identical, so that the very same disease, or branch, has the same color in all 3 graphics;
	\item To ease the readability of the treemaps, we limited them to 3 levels of hierarchy; however we enable the user to select the lowest level to be displayed since it happens sometimes that the default one is not detailed enough;
	\item Even if the process deals with the whole corpus of documents, it is possible to produce the above results for a selection of data sources, and with activating the additional filter;
	\item It is also possible to use the selection of a disease or a branch to go back to the original documents out of which it has been extracted.
\end{itemize}

The dashboard is available freely on the web. As the URL might change over time, a reasonable entry point is the following repository \url{https://github.com/fran6w/vidar-19}. It has been developed with the \emph{dash} framework provided by \emph{plotly} \cite{plotly}.

As the dashboard is regularly updated, the graphics obtained from the application may differ from the ones shown in this paper.

\section{Discussion and further work}

As explained in a recent web publication in \emph{Science Biology}: ``Despite the more than 1,000 papers now spilling into journals and onto preprint servers every week, a clear picture is elusive, as the virus acts like no microbe humanity has ever seen. Without larger, prospective controlled studies that are only now being launched, scientists must pull information from small studies and case reports, often published at warp speed and not yet peer reviewed'' \cite{science}.

We think that this project might be a mean to help the situation described above.

\subsection{Discussion}\label{discussion}

\begin{figure}[h]
	\includegraphics[width=6.3in]{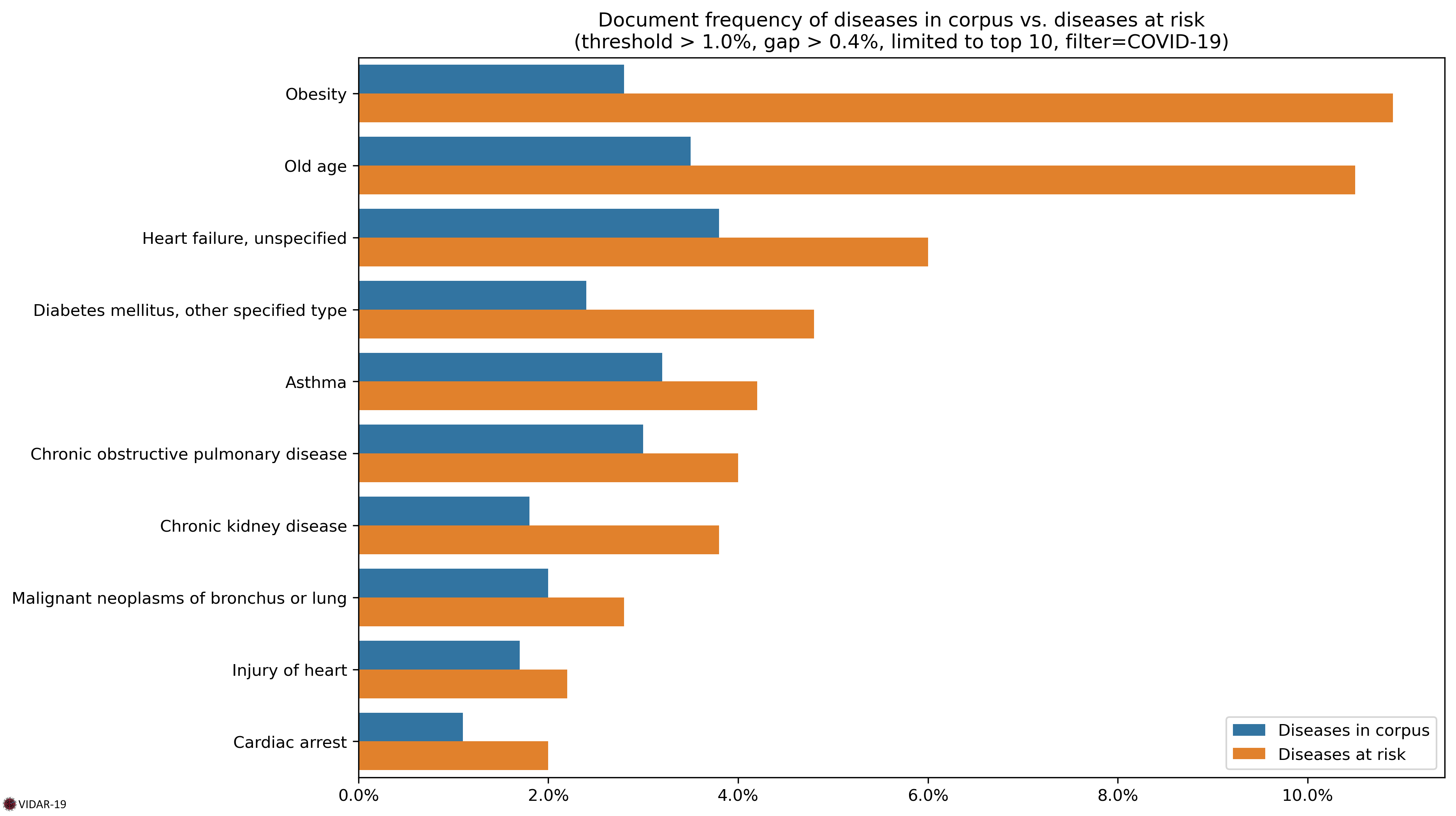}
	\caption{Document frequency of diseases in the corpus vs. diseases at risk}
	\label{gaps}
\end{figure}

If we look precisely to the quantified chart \ref{gaps} presented in section \ref{frequencies}, we can notice that it produces a list of diseases at risk for COVID-19 which shares some of the diseases stressed by the CDC: e.g., people 65 years and older, chronic lung disease, serious heart conditions, immunocompromised, severe obesity, diabetes, chronic kidney disease, liver disease \cite{cdcinfo}.

Our list is also close to the results presented in a recent paper by a team of the NYU Grossman School, which studied the factors associated with hospitalization and critical illness among 4,103 patients with COVID-19 disease: ``We find particularly strong associations of older age, obesity, heart failure and chronic kidney disease with hospitalization risk'' \cite{nyu}.

Moreover, dealing with obesity -- one of the top ranked diseases in our computations -- it is compatible with this last study which puts an emphasis on this condition: ``In this regard it is notable that the chronic condition with the strongest association with critical illness was obesity, with a substantially higher odds ratio than any cardiovascular or pulmonary disease''.

We think that theses early findings validate our approach.

A point is that the system does not take into account the number of cases reported in the articles, and this might introduce a bias into the results. Our position is that this number is implicitly contained within the number of articles, since a high-risk factor will likely be more frequently reported than a low-risk one.

\subsection{Further work}

In this paper, we have presented, VIDAR-19, a piece of software able to extract automatically diseases at risk in a corpus of documents, which has been specifically applied to the COVID-19 literature. Beyond the coronavirus disease, we think that this project has a much broader applications:

\begin{itemize}
	\item Processing any corpus of scholarly articles dealing with other pathologies in order to visualize and to explore the diseases at risk related to this pathology.
	\item Processing any corpus but looking for disease contexts other than risk factors, e.g. genetic variations, molecules therapy.
	\item Processing other corpus and extending to other classification hierarchies: e.g., geographic taxonomies, phylogenetic nomenclatures.
\end{itemize}

Beyond its dramatic aftermath, we think that the COVID-19 is an opportunity to develop methods and tools that might be useful in other cases.

\begin{flushleft}
	\textbf{\Large Acknowledgments}
\end{flushleft}

We would like to thank Prof. Jean-François Perrot and Frantz Vichot for their useful comments on this project.

\bibliography{vidar}
\bibliographystyle{hieeetr}

\clearpage

\begin{figure}[h]
	\centering
	\includegraphics[width=5.5in]{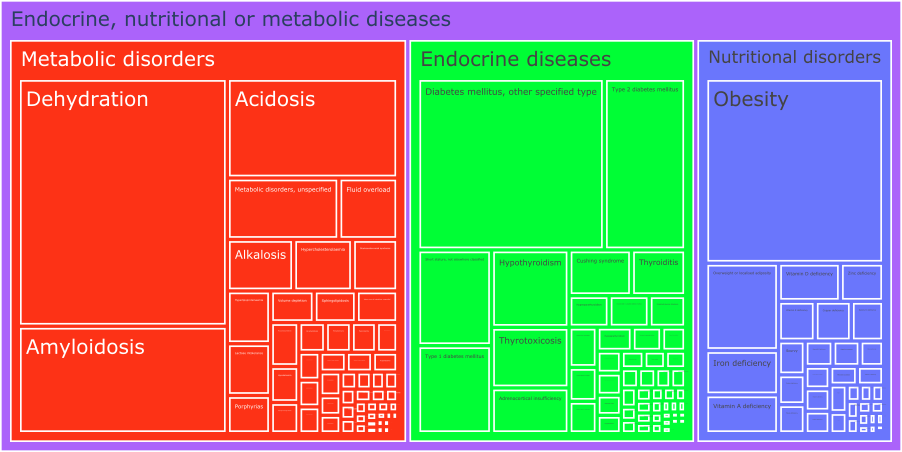}
	\caption{Occurrences of diseases in the corpus (\emph{Endocrine, nutritional or metabolic diseases} branch)}
	\label{treemap1obesity}
	\vspace{0.5in}
	\includegraphics[width=5.5in]{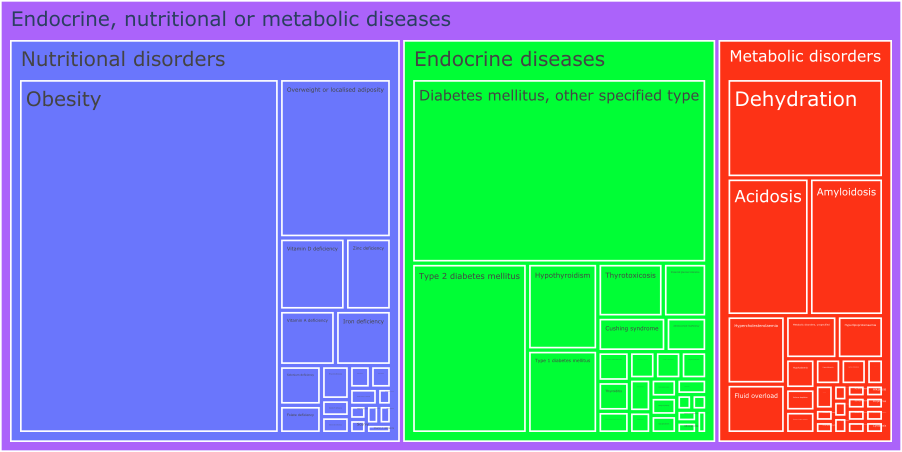}
	\caption{Occurrences of diseases at risk (\emph{Endocrine, nutritional or metabolic diseases} branch)}
	\label{treemap2obesity}
\end{figure}

\end{document}